\documentclass[epj,portrait]{svjour} 
\usepackage{graphicx,amssymb,color}
\usepackage{dcolumn}
\graphicspath{{../Figures-EPS/}}
\newcommand{\dgs}{$^\circ$}
\newcommand{\infnpd}{1}
\newcommand{\unipd}{2}
\newcommand{\infnge}{3}
\newcommand{\dresden}{4}
\newcommand{\infnmi}{5}
\newcommand{\atomki}{6}
\newcommand{\lngs}{7}
\newcommand{\infnto}{8}
\newcommand{\unina}{9}
\newcommand{\bochum}{10}
\newcommand{\teramo}{11}
\newcommand{\caserta}{12}

\begin{document}

\title{Ultra-sensitive in-beam $\gamma$-ray spectroscopy
for nuclear astrophysics at LUNA}
\author{
	A.\,Caciolli \inst{\infnpd,\unipd} \and
	L.\,Agostino \inst{\infnge} \and
	D.\,Bemmerer \inst{\dresden} \thanks{e-mail: d.bemmerer@fzd.de} \and
	R.\,Bonetti \inst{\infnmi} \thanks{Deceased.} \and
	C.\,Broggini \inst{\infnpd} \and
	F.\,Confortola \inst{\infnge} \and
	P.\,Corvisiero \inst{\infnge} \and
	H.\,Costantini \inst{\infnge} \and
	Z.\,Elekes \inst{\atomki} \and
	A.\,Formicola \inst{\lngs} \and
	Zs.\,F\"ul\"op \inst{\atomki} \and
	G.\,Gervino \inst{\infnto} \and
	A.\,Guglielmetti \inst{\infnmi} \and
	C.\,Gustavino \inst{\lngs} \and
	Gy.\,Gy\"urky \inst{\atomki} \and
	G.\,Imbriani \inst{\unina} \and
	M.\,Junker \inst{\lngs} \and
	M.\,Laubenstein \inst{\lngs} \and
	A.\,Lemut \inst{\infnge} \and
	B.\,Limata \inst{\unina} \and
	M.\,Marta \inst{\dresden} \and
	C.\,Mazzocchi \inst{\infnmi} \and
	R.\,Menegazzo \inst{\infnpd} \and
	P.\,Prati \inst{\infnge} \and
	V.\,Roca \inst{\unina} \and
	C.\,Rolfs \inst{\bochum} \and
	C.\,Rossi Alvarez \inst{\infnpd} \and
	E.\,Somorjai \inst{\atomki} \and
	O.\,Straniero \inst{\teramo} \and
	F.\,Strieder \inst{\bochum} \and
	F.\,Terrasi \inst{\caserta} \and
	H.P.\,Trautvetter \inst{\bochum} \\
	(LUNA collaboration)
	}
\institute{
	INFN Sezione di Padova, Padova, Italy 
	\and 
	Dipartimento di Fisica, Universit\`a di Padova, Padova, Italy 
	\and
	Dipartimento di Fisica, Universit\`a di Genova, and INFN Sezione di Genova, Genova, Italy 
	\and 
	Forschungszentrum Dresden-Rossendorf, Dresden, Germany 
	\and 
	Istituto di Fisica Generale Applicata, Universit\`a di Milano, and INFN Sezione di Milano, Milano, Italy 
	\and 
	ATOMKI, Debrecen, Hungary 
	\and
	INFN, Laboratori Nazionali del Gran Sasso, Assergi, Italy 
	\and 
	Dipartimento di Fisica Sperimentale, Universit\`a di Torino, and INFN Sezione di Torino, Torino, Italy 
	\and
	Dipartimento di Scienze Fisiche, Universit\`a di Napoli "Federico II", and INFN Sezione di Napoli, Napoli, Italy 
	\and
	Institut f\"ur Experimentalphysik III, Ruhr-Universit\"at Bochum, Bochum, Germany 
	\and
	Osservatorio Astronomico di Collurania, Teramo, and INFN Sezione di Napoli, Napoli, Italy 
	\and
	Seconda Universit\`a di Napoli, Caserta, and INFN Sezione di Napoli, Napoli, Italy 
	}
%
\date{As accepted by Eur. Phys. J. A, 15 December 2008}

\abstract{
Ultra-sensitive in-beam $\gamma$-ray spectroscopy studies for nuclear astrophysics are performed at the LUNA (Laboratory for Underground Nuclear Astrophysics) 400\,kV accelerator, deep underground in Italy's Gran Sasso laboratory. By virtue of a specially constructed passive shield, the laboratory $\gamma$-ray background for $E_\gamma$ $<$ 3\,MeV at LUNA has been reduced to levels comparable to those experienced in dedicated offline underground $\gamma$-counting setups. The $\gamma$-ray background induced by an incident $\alpha$-beam has been studied. The data are used to evaluate the feasibility of sensitive in-beam experiments at LUNA and, by extension, at similar proposed facilities.
}
\authorrunning{A.\,Caciolli {\it et al.} (LUNA collab.)}

\maketitle

\section{Introduction}
\label{Introduction}

The Laboratory for Underground Nuclear Astrophysics (LUNA) \cite{Greife94-NIMA} in Italy's Gran Sasso national laboratory \linebreak (LNGS, Laboratori Nazionali del Gran Sasso) is the first and, to date, only accelerator facility running deep underground. It is dedicated to the study of astrophysically relevant nuclear reactions directly at or near the energies of astrophysical relevance. The LNGS rock overburden of 3800 meters water equivalent attenuates the flux of cosmic-ray induced muons by six orders of magnitude with respect to the Earth's surface \cite{MACRO90-PLB}. The neutron flux at LNGS is three orders of magnitude lower than at the Earth's surface \cite{Belli89-NCA}. 

Motivated by the successful study of several astrophysically relevant nuclear reactions at LUNA, new underground accelerators are proposed e.g. at LNGS \cite{NuPECC05-Roadmap}, at the planned DUSEL facility in the United States \cite{Haxton07-NIMA,DOE08-arxiv}, at Boulby mine in the United Kingdom \cite{Strieder08-NPA3}, and at several possible sites in Romania \cite{Bordeanu08-NPA3}. Like the existing LUNA facility, these new proposals are driven by the need for precise data for astrophysical applications. 

However, more general analysis techniques have already benefited from a great increase in sensitivity owing to the introduction of offline underground $\gamma$-counting with well-shielded high-purity germanium (HPGe) detectors \cite{Arpesella96-Apradiso,Laubenstein04-Apradiso}. Therefore it is conceivable that also in-beam analysis techniques involving $\gamma$-ray detection \cite{Johansson95} may benefit from the laboratory background suppression achieved by going underground. 

\begin{figure*}[bt]
  \center{\includegraphics[width=0.8\textwidth]{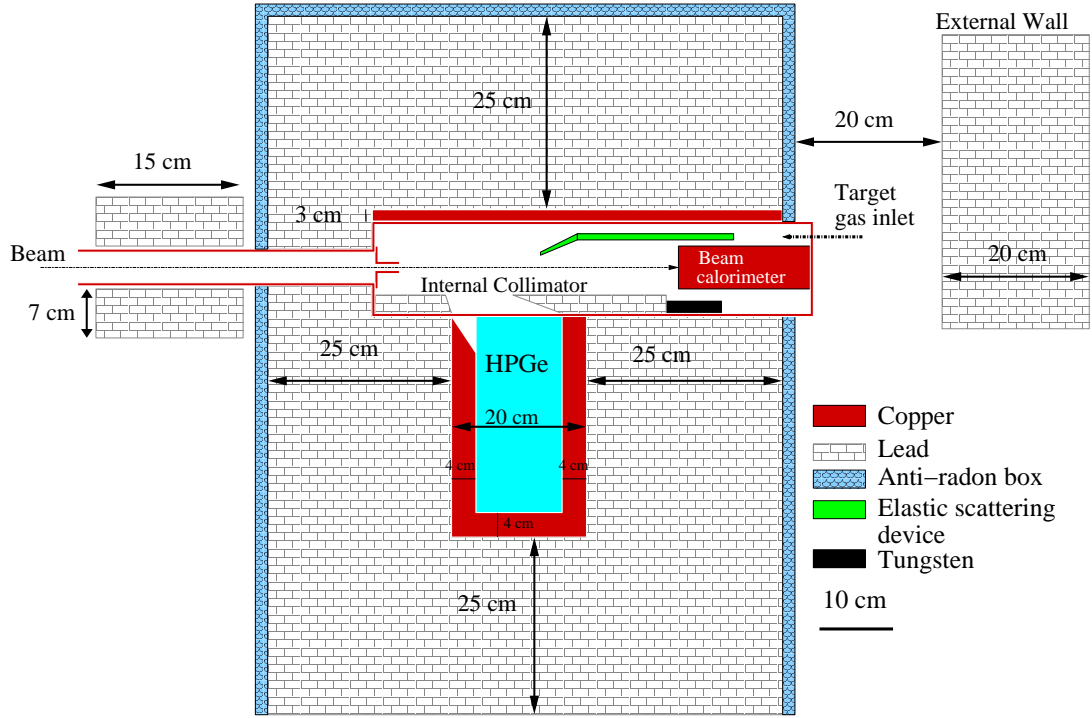}}
  \caption{Schematic view of experimental setup C (complete shielding). }
  \label{fig:Setup}
\end{figure*}

In a previous work \cite{Bemmerer05-EPJA}, the feasibility of radiative capture experiments at LUNA has been investigated for $\gamma$-ray energies above 3\,MeV, and the $\gamma$-ray background induced by a proton beam has been localized using the Doppler shift. LUNA experiments where the analysis concentrates on $\gamma$-rays with $E_\gamma$ $>$ 3\,MeV are special in that it is not necessary to strongly shield the detector against laboratory $\gamma$-ray background, simply because this background is negligible at LUNA \cite{Bemmerer05-EPJA}, due to the reduced cosmic ray flux. When $\gamma$-rays with $E_\gamma$ $<$ 3\,MeV are to be detected, however, the picture changes. For these low $\gamma$-ray energies, natural radioisotopes present at the LUNA site dominate the background, and a sophisticated shielding of setup and detector is required. 

The $\gamma$-rays with $E_\gamma$ $>$ 3\,MeV discussed in the previous study \cite{Bemmerer05-EPJA} are characteristic of radiative capture reactions with $Q$-values also above 3\,MeV. In recent years, several such reactions have been studied at LUNA or are presently under study: 
\begin{itemize}
\item the $^{2}$H(p,$\gamma$)$^{3}$He reaction \cite{Casella02-NPA}, $Q$-value 5.493\,MeV, important for hydrogen burning by the proton-proton chain in the Sun \cite{Bahcall05-ApJ}, 
\item the $^{14}$N(p,$\gamma$)$^{15}$O reaction \cite{Formicola04-PLB,Imbriani05-EPJA,Lemut06-PLB,Bemmerer06-NPA,Marta08-PRC}, $Q$-value 7.297\,MeV, bottleneck of the CNO cycle, important for solar neutrinos \cite{Haxton08-ApJ}, globular cluster ages \cite{Imbriani04-AA}, and hydrogen shell burning in asymptotic giant branch stars \cite{Herwig04-ApJ,Herwig06-PRC}, 
\item the $^{25}$Mg(p,$\gamma$)$^{26}$Al reaction \cite{Formicola08-NPA3}, $Q$-value 6.306\,MeV, controlling the nucleosynthesis of radioactive $^{26}$Al, a tracer of live nucleosynthesis \cite{Diehl06-Nature}, and 
\item the $^{15}$N(p,$\gamma$)$^{16}$O reaction, $Q$-value 12.127\,MeV, important in nova nucleosynthesis \cite{Iliadis02-ApJSS}.
\end{itemize}
Recently, the technique of underground in-beam $\gamma$-spectro\-metry has been extended to radiative capture reactions with $Q$-values below 3\,MeV: 
\begin{itemize}
\item The $^{3}$He($\alpha$,$\gamma$)$^{7}$Be reaction has been studied at LUNA, both by activation \cite{Bemmerer06-PRL,Gyurky07-PRC,Confortola07-PRC} and by in-beam $\gamma$-spectro\-metry \cite{Confortola07-PRC,Costantini08-NPA}. This reaction has a $Q$-value of 1.586\,MeV. It controls the flux of $^7$Be and $^8$B neutrinos from the Sun \cite{Bahcall05-ApJ,Haxton08-ApJ} and the production of $^7$Li in big-bang nucleosynthesis \cite{Serpico04-JCAP}.
\item A study of the $^{2}$H($\alpha$,$\gamma$)$^{6}$Li reaction is planned at LUNA. This reaction has a $Q$-value of 1.474\,MeV and is important for big-bang nucleosynthesis \cite{Serpico04-JCAP}.
\end{itemize}
Further astrophysically important reactions with low $Q$-value that merit study are, for example, 
\begin{itemize}
\item the $^{12}$C(p,$\gamma$)$^{13}$N reaction \cite{Rolfs74-NPA-12C}, $Q$-value 1.943\,MeV, important for pre-equilibrium CNO burning \cite{Haxton08-ApJ},
\item the $^{12}$C($^{12}$C,$\alpha$)$^{20}$Ne ($Q$-value 4.617\,MeV, main $\gamma$-ray energy 1634\,keV) and $^{12}$C($^{12}$C,p)$^{23}$Na ($Q$-value \linebreak 2.241\,MeV, main $\gamma$-ray energy 440\,keV) reactions \cite{Spillane07-PRL}, important for carbon burning in massive stars \cite{Iliadis07}, and
\item the $^{24}$Mg(p,$\gamma$)$^{25}$Al reaction \cite{Powell99-NPA}, $Q$-value 2.272\,MeV, important for hydrogen burning in massive stars \cite{Iliadis07}. 
\end{itemize}

The aim of the present work is to facilitate the underground study of radiative capture reactions for nuclear astrophysics. It concentrates on reactions with low $Q$-value or low energy of the emitted $\gamma$-rays, extending the previous study \cite{Bemmerer05-EPJA} to $\gamma$-ray energies below 3\,MeV. The present considerations apply not only at LUNA, but can be extended to other potential underground accelerator sites \cite{Haxton07-NIMA,DOE08-arxiv,Strieder08-NPA3,Bordeanu08-NPA3}.

In addition, the previous study of proton-beam-induced $\gamma$-ray background \cite{Bemmerer05-EPJA} is taken one step further here, studying the $\gamma$-ray background induced by an intensive $\alpha$-beam. 

The present work is organized as follows. In section 2, the experimental setup is described. Section 3 shows the laboratory $\gamma$-ray background observed in several stages of completion of the setup. Section 4 reports on in-beam $\gamma$-ray background studies with an intensive $\alpha$-beam. In section 5, the background data are used to evaluate the feasibility of in-beam $\gamma$-spectroscopic experiments deep underground.  

\section{Experimental setup}

The setup is sited at the LUNA2 400~kV accelerator \cite{Formicola03-NIMA} facility. It consists of a windowless, differentially pumped gas target and a shielded HPGe detector described below. The construction of this ultra-sensitive setup was necessary for the LUNA experiment on the $^3$He($\alpha$,$\gamma$)$^7$Be reaction at unprecedented low energies \cite{Bemmerer06-PRL,Gyurky07-PRC,Confortola07-PRC,Costantini08-NPA}. 

The HPGe detector is a Canberra ultra-low background p-type coaxial detector with 137\% relative efficiency. The endcap of the detector is made of low-background copper, and the cryostat is connected to the crystal by a 25\,cm long cold finger. The crystal is oriented at 90\dgs\ with respect to the cold finger, so that the direct line of sight from the cryostat to the crystal can be shielded by a 25\,cm thick layer of lead.

The ion beam from the LUNA2 accelerator first passes a disk-shaped watercooled collimator with 7\,mm inner diameter, and then it enters the gas target chamber. The target chamber (fig.~\ref{fig:Setup}), made of oxygen free high conductivity (OFHC) copper, is 60\,cm long and has 12\,cm by 11\,cm area. The ion beam is stopped, inside the target chamber, on a copper disk that serves as the hot side of a beam calorimeter with constant temperature gradient \cite{Casella02-NIMA}. 

\begin{figure}[tb]
  \includegraphics[width=\columnwidth]{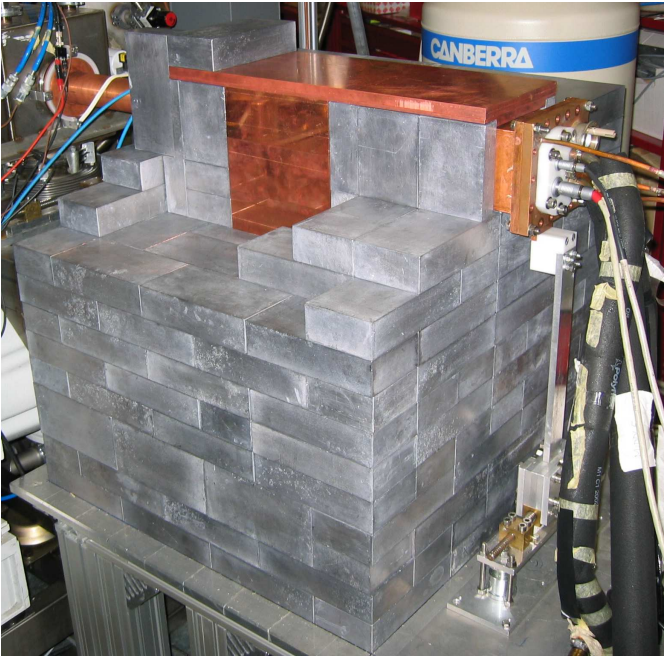}%
  \caption{Photo taken during the construction of the shield. The lower half of the lead shield is already complete, as well as the copper shield for the detector and the copper plate above the target chamber. The back plate and connectors of the beam calorimeter are visible in the upper right corner. The copper tube for the beam inlet is seen in the upper left corner. The upper half of the lead shield was not yet installed when the photo was taken.}
  \label{fig:Menegazzo-182}
\end{figure}

The shielding consists of several layers and surrounds detector and target chamber, excepting two holes for letting in the ion beam and for the beam calorimeter. It is designed in such a way that the germanium crystal of the detector is typically shielded by 4\,cm copper and 25\,cm lead. The innermost shielding layer surrounding the detector is made of OFHC copper bricks machined so that the detector fits inside with only 1-2\,mm of space left free. A 3\,cm thick OFHC copper plate above the target chamber carries the weight of the upper half of the lead shield (fig.~\ref{fig:Menegazzo-182}). The remainder of the shield is made of lead bricks with low $^{210}$Pb content (25 Bq/kg $^{210}$Pb, supplied by JL-Goslar, Germany) and is 25\,cm thick. The lead bricks have been cleaned with citric acid prior to mounting, in order to remove accumulated dust and surface oxidation. In order to avoid $\gamma$-rays from the decay of radon daughters, the setup is enclosed in a plexiglass anti-radon box that is flushed with the nitrogen gas evaporating from the HPGe detector's dewar. The gas volume inside the anti-radon box is approximately 4\,liters.

Outside the anti-radon box, a 15\,cm thick wall of the aforementioned low-background lead is placed upstream of the target chamber. In addition, a 20\,cm thick wall of the same lead is placed behind the end of the calorimeter (fig.~\ref{fig:Setup}). Inside the target chamber, the $\gamma$-rays emitted within the gas target are collimated by 3\,cm thick trapezoidal-shaped lead bricks that also serve as additional shield. An elastic scattering device for studies of effective target gas density and gas contaminations is included inside the chamber \cite{Marta06-NIMA}. In order to limit possible $\gamma$-emissions, the elastic scattering device has been made of Delrin. 

For the purpose of the present study, three experimental configurations called setups A, B, and C are considered. Setups A, B, and C are all sited in the LUNA2 accelerator room deep underground.

\begin{enumerate}
\item[A.] HPGe detector without any shield.  
\item[B.] HPGe detector with complete shield except for
\begin{itemize}
\item inner trapezoidal lead collimator,
\item 20\,cm lead wall behind the calorimeter, and
\item anti-radon box.
\end{itemize}
\item[C.] HPGe detector with complete shield (fig.~\ref{fig:Setup}).  
\end{enumerate}

For comparison, also a fourth experimental configuration is considered, here called setup LLL: A HPGe detector of similar size (125\% relative efficiency) and equal geometry to the present one. It is shielded with 25\,cm low-background lead including an inner lining of quasi $^{210}$Pb-free lead from a sunken Roman ship, and it has a highly efficient anti-radon box \cite{Arpesella96-Apradiso}. This detector is placed outside the LUNA2 accelerator room, in the LNGS low-background laboratory (LLL) \cite{Laubenstein04-Apradiso}. Setup LLL is dedicated to measurements of extremely low $\gamma$-activities, as opposed to the in-beam setups A-C. As a consequence, no entrance pipe for the ion beam has been provided in setup LLL, improving the shielding.

Setup C has been used for the in-beam $\gamma$-spectroscopic part of the LUNA $^3$He($\alpha$,$\gamma$)$^7$Be study \cite{Confortola07-PRC,Costantini08-NPA}. Setup LLL has been used for part of the $^7$Be-activity counting in that same study \cite{Bemmerer06-PRL,Gyurky07-PRC,Confortola07-PRC,Costantini08-NPA}. 

\section{Laboratory $\gamma$-ray background studies for $E_\gamma$ $<$ 3\,MeV}

The laboratory $\gamma$-ray background has been studied for setups A, B, and C, with running times of several days without ion beam for each setup. For comparison, also the spectrum taken with an inert sample ($^4$He+$^4$He irradiated OFHC copper \cite{Bemmerer06-PRL}) on detector LLL is shown.

\begin{figure}[tb]
  \includegraphics[width=\columnwidth]{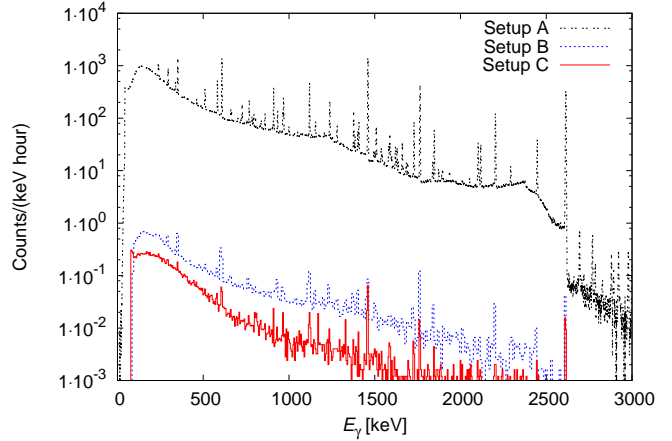}%
  \caption{Laboratory $\gamma$-ray background spectra for setups A (black dashed line), B (blue dotted line), and C (red full line). See table~\ref{table:LineBackground} for the counting rate of selected lines and table~\ref{table:LineList} for the assignment of the lines evident in spectrum C.}
  \label{fig:LabBG-Spectra1}
\end{figure}

Comparing the unshielded setup A with the shielded setup B (fig.~\ref{fig:LabBG-Spectra1}), a reduction of three orders of magnitude in the $\gamma$-ray continuum below 2615\,keV is observed, and the summing lines above the 2615\,keV $^{208}$Tl line are no longer evident. In addition, the counting rate for the most important single $\gamma$-lines is reduced by three orders of magnitude or more (table~\ref{table:LineBackground}). 

\begin{table*}[tb] 
 \centering
\begin{tabular}{|l|c|c|c|c|c|c|c|}
\hline
$E_\gamma$ [keV] & 511 & 609 & 1120 & 1461 & 1730 & 1764 & 2615 \\
\hline
Setup A, offline  & 762$\pm$4 & 3729$\pm$4 & 1278$\pm$3 & 4870$\pm$4 & 246$\pm$2 & 1350$\pm$2 & 1325$\pm$2 \\
Setup B, offline & 0.60$\pm$0.13 & 3.9$\pm$0.2 & 1.33$\pm$0.12 & 0.93$\pm$0.11 & 0.26$\pm$0.06 & 1.18$\pm$0.10 & 0.42$\pm$0.06 \\
Setup B, $\alpha$-beam  & 1.74$\pm$0.32 & 4.6$\pm$0.4 & 1.9$\pm$0.3 & 0.51$\pm$0.17 & 0.32$\pm$0.11 & 1.06$\pm$0.19 & 0.55$\pm$0.11 \\
Setup C, offline & 0.09$\pm$0.04 & 0.30$\pm$0.04 & 0.15$\pm$0.02 & 0.42$\pm$0.03 & 0.038$\pm$0.010 & 0.098$\pm$0.014 & 0.12$\pm$0.02 \\
Setup C, $\alpha$-beam & 0.32$\pm$0.13 & 0.31$\pm$0.10 & $<$0.08 & 0.37$\pm$0.08 & $<$0.04 & 0.08$\pm$0.05 & 0.06$\pm$0.03 \\
Setup LLL, offline & $<$0.04 & 0.055$\pm$0.018 & 0.013$\pm$0.008 & 0.098$\pm$0.018 & $<$0.011 & 0.012$\pm$0.007 & 0.016$\pm$0.005 \\
\hline
\end{tabular}
 \caption{Counting rate in counts/hour for selected $\gamma$-lines. The laboratory background (offline) runs in setups A, B, and C are discussed in section 3. The in-beam runs in setups B and C are discussed in section 4. For comparison, the corresponding numbers are also given for an inert sample counted in setup LLL at the LNGS low level laboratory. Upper limits, where applicable, are given for 2$\sigma$ confidence level.}
 \label{table:LineBackground}
\end{table*}
\begin{table*}[tb] 
\begin{center}
\begin{tabular}{|l|c|c|c|c|c|}
\hline
Reaction & $^{12}$C($^{12}$C,p)$^{23}$Na & $^{2}$H($\alpha$,$\gamma$)$^{6}$Li & $^{12}$C($^{12}$C,$\alpha$)$^{20}$Ne & $^{3}$He($\alpha$,$\gamma$)$^{7}$Be & $^{12}$C(p,$\gamma$)$^{13}$N \\
$\gamma$-ray ROI [keV] & 425-455 & 1555-1585 & 1619-1649 & 1731-1761  & 2004-2034 \\
\hline
Setup A, offline & (2.016$\pm$0.002)$\cdot$10$^{2}$ & (1.585$\pm$0.004)$\cdot$10$^{1}$ & (1.481$\pm$0.004)$\cdot$10$^{1}$ & (1.332$\pm$0.004)$\cdot$10$^{1}$ & (5.73$\pm$0.02)$\cdot$10$^{0}$ \\
Setup B, offline & (1.70$\pm$0.07)$\cdot$10$^{-1}$ & (1.3$\pm$0.2)$\cdot$10$^{-2}$ & (1.1$\pm$0.2)$\cdot$10$^{-2}$ & (1.1$\pm$0.2)$\cdot$10$^{-2}$ & (6.7$\pm$1.4)$\cdot$10$^{-3}$ \\
Setup B, $\alpha$-beam & (1.87$\pm$0.12)$\cdot$10$^{-1}$ & (2.1$\pm$0.4)$\cdot$10$^{-2}$ & (1.1$\pm$0.3)$\cdot$10$^{-2}$ & (6$\pm$2)$\cdot$10$^{-3}$ & (4$\pm$2)$\cdot$10$^{-3}$ \\
Setup C, offline & (7.2$\pm$0.2)$\cdot$10$^{-2}$ & (1.5$\pm$0.3)$\cdot$10$^{-3}$ & (1.4$\pm$0.3)$\cdot$10$^{-3}$ & (1.1$\pm$0.3)$\cdot$10$^{-3}$ & (9$\pm$3)$\cdot$10$^{-4}$ \\
Setup C, $\alpha$-beam  & (7.8$\pm$0.6)$\cdot$10$^{-2}$ & (2.7$\pm$1.2)$\cdot$10$^{-3}$ & (1.1$\pm$0.8)$\cdot$10$^{-3}$ & (2.2$\pm$1.1)$\cdot$10$^{-3}$ & (1.1$\pm$0.8)$\cdot$10$^{-3}$ \\
\hline
\end{tabular}
\end{center}
 \caption{Same as table~\ref{table:LineBackground}, but instead of $\gamma$-line counting rates, the continuum background rate in counts/(keV hour) is given. The reactions mentioned in the first line of the table are discussed in section 5.}
 \label{table:ContBackground}
\end{table*}

Improving the shielding from setup B to the final setup C yields up to another order of magnitude suppression in the $\gamma$-continuum below 2615\,keV. The $^{40}$K (1461\,keV) and $^{208}$Tl (2615\,keV) lines are reduced by a factor 2 and 3, respectively. The counting rates of these two lines are dominated by $\gamma$-emitters outside the setup: in construction materials for $^{40}$K, in the walls of the LNGS tunnel for the Thorium daughter $^{208}$Tl. These sources are already well shielded by setup B. The 15\,cm thick lead wall behind the calorimeter and the internal lead collimator improve the effective shielding thickness for a limited solid angle; hence the limited improvement of only a factor 2-3.

The counting rates in the 609, 1120, 1730, and 1764\,keV lines (all assigned to $^{214}$Bi in the present case) are improved by about a factor 10 from setup B to setup C. This is due to the operation of the anti-radon box that reduces the amount of $^{222}$Rn ($t_{\rm 1/2}$ = 3.8 d, progenitor of $^{214}$Bi) present in the remaining air pockets near the detector. 

A similar reduction is evident for the 511\,keV annihilation line in setup C; it is just barely significant at 2$\sigma$ level. Because of the deep-underground location, muon-induced pair production and decay of stopped $\mu^+$ \cite{Heusser93-NIMB} only contribute negligibly to the counting rate in that line. Different from $\gamma$-ray spectroscopy systems at the surface of the Earth, in the present case this line can therefore be effectively attenuated by passive shielding against $\beta^+$ emitters, and by eliminating $\beta^+$-emitting contaminations.

The remaining $\gamma$-lines evident in spectrum C (fig.~\ref{fig:LabBG-Spectra1}) can all be traced back to natural radionuclides present in the laboratory, detector, or radon gas (table~\ref{table:LineList}). No neutron-induced (n,$\gamma$), (n,n'$\gamma$), or activation lines \cite{Heusser93-NIMB} can be identified in spectrum C after 21\,days counting time. 

\begin{figure}[tb]
  \includegraphics[width=\columnwidth]{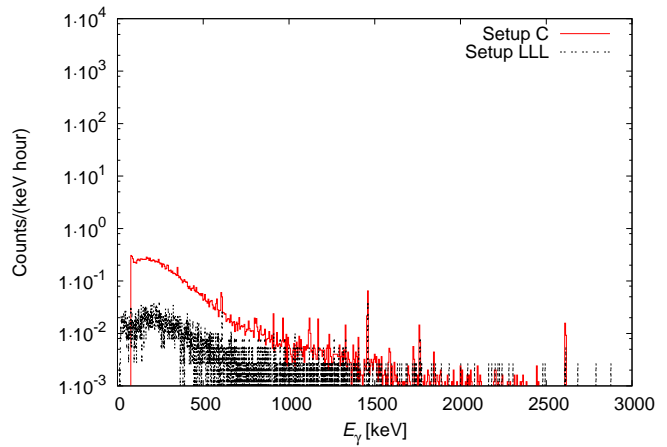}%
  \caption{Laboratory $\gamma$-ray background spectra, setup C (red full line, same as in fig.~\ref{fig:LabBG-Spectra1}), and inert sample counted in setup LLL (black dashed line). See table~\ref{table:LineBackground} for the counting rate of selected lines and table~\ref{table:LineList} for the assignment of the lines evident in spectrum C.}
  \label{fig:LabBG-Spectra2}
\end{figure}

In order to judge the quality of the background suppression in spectrum C, this spectrum is compared with an inert sample \cite{Bemmerer06-PRL} counted in setup LLL (fig.~\ref{fig:LabBG-Spectra2}). The spectra have not been corrected for the slightly different size of the crystals: 137\% in setups A-C and 125\% in setup LLL. The $\gamma$-continuum below 1162\,keV ($Q$-value of the $\beta^-$-decay of $^{210}$Bi, daughter of $^{210}$Pb) is up to one order of magnitude higher in setup C when compared to setup LLL. Some of this continuum stems from bremsstrahlung emitted by electrons created in $^{210}$Bi $\beta^-$ decay. The modern low-background lead used in setup C still has more $^{210}$Pb than the Roman ship lead used for the inner shield lining in setup LLL.  

The higher counting rates in the $^{40}$K (1461\,keV), $^{208}$Tl (2615\,keV), and radon (609, 1120, 1730, and 1764\,keV) lines in setup C with respect to setup LLL, with a concomitant increase in the Compton continuum, are ascribed to the inevitable opening for the beam pipe, which leads to a small window of not optimally shielded solid angle.

\begin{table}[tb] 
 \centering
\begin{tabular}{|r|l|l|}
\hline
$E_\gamma$ [keV] & Nuclide & Source \\
\hline
352 & $^{214}$Pb & Radon gas \\
511 & e$^+$ e$^-$ & $\beta^+$ emitters\\
583 & $^{208}$Tl & Thorium chain, tunnel walls \\
609 & $^{214}$Bi & Radon gas \\
768 & $^{214}$Bi & Radon gas \\
911 & $^{228}$Ac & Thorium chain, tunnel walls \\
965 & $^{228}$Ac & Thorium chain, tunnel walls \\
969 & $^{228}$Ac & Thorium chain, tunnel walls \\
1120 & $^{214}$Bi & Radon gas \\
1173 & $^{60}$Co & Detector contamination \\
1238 & $^{214}$Bi & Radon gas \\
1333 & $^{60}$Co & Detector contamination \\
1408 & $^{214}$Bi & Radon gas \\
1461 & $^{40}$K & Potassium in construction materials\\
1588 & $^{228}$Ac & Thorium chain, tunnel walls \\
1730 & $^{214}$Bi & Radon gas \\
1764 & $^{214}$Bi & Radon gas \\
1847 & $^{214}$Bi & Radon gas \\
2615 & $^{208}$Tl & Thorium chain, tunnel walls\\
\hline
\end{tabular}
 \caption{List of $\gamma$-lines evident at 2$\sigma$ level in the laboratory background of setup C (fig.~\ref{fig:LabBG-Spectra1}).}
 \label{table:LineList}
\end{table}

\section{$\gamma$-ray background induced by the $\alpha$-beam}

As a next step, two experiments with $\alpha$-beam have been performed.

\begin{figure}[tb]
  \includegraphics[width=\columnwidth]{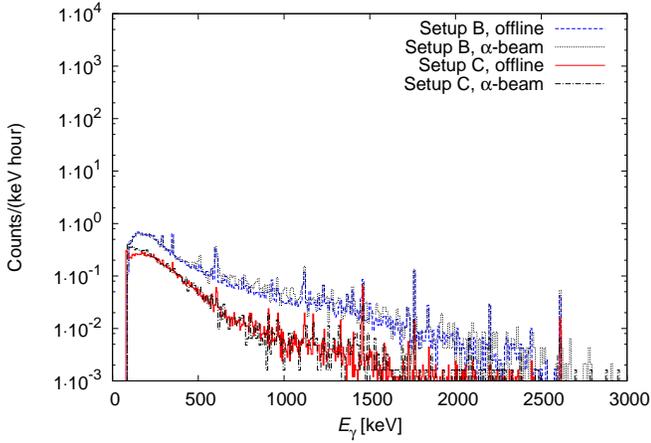}
  \caption{Comparison of offline and in-beam $\gamma$-spectra: Setup B, offline (blue dashed line) and with $\alpha$-beam (black dotted line). Setup C, offline (red full line) and with $\alpha$-beam (black dot-dashed line). }
  \label{fig:InBeam-Spectrum}
\end{figure}

In the first in-beam experiment, the beamstop in setup B has been bombarded for 47 hours with a $^4$He$^+$-beam of $E_\alpha$ = 350\,keV and 110\,$\mu$A intensity from the LUNA2 accelerator. During this experiment, the gas target setup was evacuated to better than 10$^{-3}$\,mbar. The in-beam spectrum shows no additional $\gamma$-lines with respect to the laboratory background (fig.~\ref{fig:InBeam-Spectrum}). 

The only difference between the in-beam and offline spectra in setup B that is significant at 2$\sigma$ level is a higher counting rate in the 511\,keV e$^+$e$^-$ annihilation line. The continuum counting rates of relevance for in-beam $\gamma$-spectroscopic studies are consistent (table~\ref{table:ContBackground}). In order to explain the 511\,keV counting rate increase, one has to assume some creation of $\beta^+$ emitters by the $\alpha$-beam. Due to the absence of other new $\gamma$-lines in the spectrum, it is impossible to assign a particular nuclide (and, by extension, a particular reaction producing that nuclide) as the supposed $\beta^+$ emitter. However, it should be noted that the in-beam 511\,keV counting rate is still 400 times lower than in the unshielded case of setup A.

The second in-beam experiment, was performed with the fully shielded setup C and a $^4$He$^+$-beam of $E_\alpha$ = 400\,keV and 240\,$\mu$A intensity for a 62 hour long irradiation \cite{Confortola07-PRC,Costantini08-NPA}. The gas target was filled with 0.7\,mbar $^4$He gas (chemical purity 99.9999\%) during this experiment. Just as in the previous case, the in-beam spectrum shows no additional $\gamma$-lines with respect to the corresponding laboratory background (fig.~\ref{fig:InBeam-Spectrum}). 

The line counting rates (table~\ref{table:LineBackground}), as well as the continuum counting rates of relevance for in-beam $\gamma$-spectro\-scopic studies  (table~\ref{table:ContBackground}) are consistent between in-beam and offline runs in setup C. Only at the lowest $\gamma$-ray energies, the in-beam continuum counting rate is slightly higher, but this difference is only significant at the 1$\sigma$ level. For the 511\,keV line, a slightly higher counting rate is observed in the in-beam spectrum, but the increase is not significant at the 2$\sigma$ level. 

Summarizing, in two exemplary configurations without gas ($p$ $<$ 0.001\,mbar) and with 0.7\,mbar $^4$He gas in the target, the $\alpha$-beam induced background was shown to be negligible when compared with the already low laboratory background. This conclusion has been verified for a number of specific nuclear reactions (table~\ref{table:ContBackground}).

\begin{figure*}[t]
  \includegraphics[width=\textwidth]{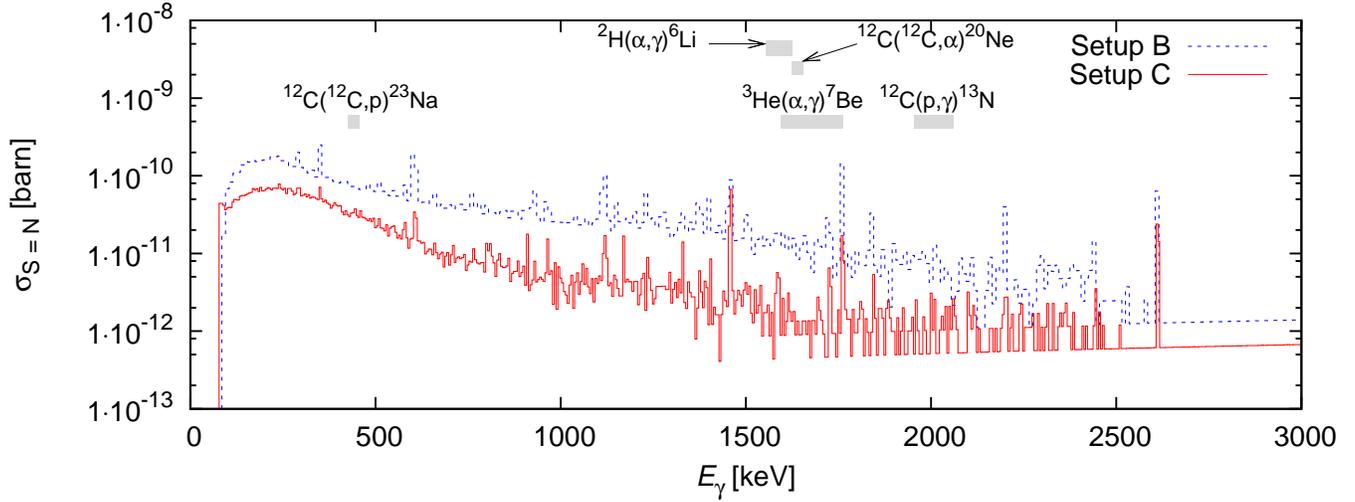}
  \caption{Cross-section $\sigma_{\rm S=N}(E_\gamma)$ defined by eq.~(\ref{eq:sigmadef}), for setups B (blue dashed line) and C (red full line). The $\gamma$-ray energies of interest for the nuclear reactions listed in tables \ref{table:ContBackground} and \ref{table:Reactions} are indicated by gray bars.}
  \label{fig:minimumsigma}
\end{figure*}

\section{Feasibility of cross-section measurements at LUNA}

In order to evaluate the feasibility of in-beam cross-section measurements, the present background data are used to calculate the hypothetical cross-section $\sigma_{\rm S=N}(E_\gamma)$ for which the expected 'signal' S would be equal to the 'noise' N, here taken as the laboratory $\gamma$-ray background in a 30\,keV wide $\gamma$-ray region of interest [$E_\gamma$ - 15\,keV; $E_\gamma$ + 15\,keV]
\begin{equation}
 \sigma_{\rm S=N}(E_\gamma) = \frac{{\rm LabBG}(E_\gamma) \cdot 30\,{\rm keV}}{\varepsilon_\gamma(E_\gamma) \cdot 6 \cdot 10^{17} \frac{\rm atoms}{\rm cm^2} \cdot 250\, \mu{\rm A} \cdot 3600\, \rm s}
\label{eq:sigmadef}
\end{equation}
Here, LabBG$(E_\gamma)$ is the laboratory background counting rate per keV and hour plotted in fig.~\ref{fig:LabBG-Spectra1}, for setup B or C, respectively. The $\gamma$-ray detection efficiency in the present geometry is $\varepsilon_\gamma(E_\gamma)$ (0.4\% at $E_\gamma$ = 1.33\,MeV). An effective target thickness of  6 $\cdot$ 10$^{17}$ atoms/cm$^2$ is assumed, corresponding e.g. in the $^3$He($\alpha$,$\gamma$)$^7$Be case to 9\,keV energy loss in the target \cite{Bemmerer06-PRL,Gyurky07-PRC}. Due to the steep decline of the Coulomb barrier penetrability with lower energy, as a rule of thumb in LUNA-type experiments an increase of the target thickness beyond 10$^{18}$ atoms/cm$^2$ does not increase the yield any further \cite{Iliadis07}. As ion beam intensity, a typical LUNA value of 250\,particle-$\mu$A is assumed, and a branching ratio of 1 for the $\gamma$-ray of interest has been supposed.

\begin{table*}[tb] 
 \centering
\begin{tabular}{|l|l|l|r|c|r|c|}
\hline
Reaction & Scenario & T$_9$ & $E_{\rm Gamow}$ & $\sigma(E_{\rm Gamow})$ & $E_\gamma$ & $\sigma_{\rm S=N}$ \\
 & & [10$^9$\,K] & [keV] & [barn] & [keV] & [barn] \\
\hline
$^{2}$H($\alpha$,$\gamma$)$^{6}$Li & Big-bang nucleosynthesis & 0.3 & 96 &  2\,$\cdot$\,10$^{-11}$ & 1570 & 1.7\,$\cdot$\,10$^{-12}$ \\
$^{3}$He($\alpha$,$\gamma$)$^{7}$Be & Big-bang nucleosynthesis & 0.3 & 160 &  1\,$\cdot$\,10$^{-8}$ & 1746 & 1.3\,$\cdot$\,10$^{-12}$ \\
 & Proton-proton chain in the Sun & 0.016 & 23 &  4\,$\cdot$\,10$^{-17}$ & 1609 & 1.8\,$\cdot$\,10$^{-12}$ \\
$^{12}$C(p,$\gamma$)$^{13}$N & CNO cycle burning in the Sun & 0.016 & 25 &  2\,$\cdot$\,10$^{-17}$ & 1968 & 1.1\,$\cdot$\,10$^{-12}$ \\
 & Hydrogen shell burning & 0.085 & 76 &  3\,$\cdot$\,10$^{-11}$ & 2019 & 1.2\,$\cdot$\,10$^{-12}$ \\
$^{12}$C($^{12}$C,p)$^{23}$Na (non-res.) & Core carbon burning & 0.5 & 1500 & 2\,$\cdot$\,10$^{-16}$ & 440 & 3\,$\cdot$\,10$^{-11}$ \\
$^{12}$C($^{12}$C,$\alpha$)$^{20}$Ne (non-res.) & Core carbon burning &  0.5 & 1500 & 2\,$\cdot$\,10$^{-16}$ & 1634 & 1.6\,$\cdot$\,10$^{-12}$\\
\hline
\end{tabular}
 \caption{Reactions of astrophysical interest discussed in the text. The relevant astrophysical scenario, typical temperature T$_9$ and corresponding Gamow energy \cite{Iliadis07} $E_{\rm Gamow}$ are also given. The expected cross-section at $E_{\rm Gamow}$ has been estimated. The relevant $\gamma$-ray energy is also  shown. For the case of setup C, the cross-section $\sigma_{\rm S=N}(E_\gamma)$ has been calculated following eq.~(\ref{eq:sigmadef}). }
 \label{table:Reactions}
\end{table*}

The laboratory background level is evaluated for a 30\,keV wide region of interest in $E_\gamma$ (fig.~\ref{fig:minimumsigma}). This approach is valid for primary $\gamma$-rays from capture into a particular level in the compound nucleus, with a target thickness equivalent to 30\,keV energy loss by the primary beam. For light target nuclei like $^3$He and $^2$H, the Doppler shift for $\gamma$-rays emitted before or behind the detector makes it necessary to maintain a 30\,keV wide region of interest even if the energy loss in the target is lower.
For secondary $\gamma$-rays, the resolution of the $\gamma$-ray from the decay of the relevant excited state in the Compound nucleus (again taking Doppler corrections into account) should be adopted instead of 30\,keV in eq.~(\ref{eq:sigmadef}). This leads to a somewhat improved sensitivity, so for secondary $\gamma$-rays the present $\sigma_{\rm S=N}(E_\gamma)$ values (fig.~\ref{fig:minimumsigma}) should be taken as a conservative upper limit.  

To put the sensitivity data in the proper astrophysical context, some representative examples for nuclear reactions are worthy to be studied afresh are mentioned here. Concerning the reactions responsible for the production of $^{6,7}$Li in big-bang nucleosynthesis ($T_9$ $\approx$ 0.3-0.9; $T_9$ stands for the temperature in 10$^9$\,K), $^{2}$H($\alpha$,$\gamma$)$^{6}$Li and \linebreak $^3$He($\alpha$,$\gamma$)$^7$Be, the present sensitivity is sufficient for an experimental study directly at the Gamow energy (table~\ref{table:Reactions}). Such a study has indeed been performed in the $^3$He($\alpha$,$\gamma$)$^7$Be case \cite{Confortola07-PRC,Costantini08-NPA}. The CNO-cycle reaction \linebreak $^{12}$C(p,$\gamma$)$^{13}$N at temperatures typical for hydrogen shell burning is an analogous case; also here a direct study is feasible.

For the temperature at the center of the Sun, $T_9$ $\approx$ 0.016, however, a study directly at the Gamow energy would be hampered by the prohibitively low cross-section for the two example reactions $^3$He($\alpha$,$\gamma$)$^7$Be and\linebreak $^{12}$C(p,$\gamma$)$^{13}$N (table~\ref{table:Reactions}).

Similar considerations apply for stable carbon burning, where temperatures of $T_9$ = 0.5-1.0 are experienced. Two of the most important carbon burning reactions, \linebreak $^{12}$C($^{12}$C,$\alpha$)$^{20}$Ne and $^{12}$C($^{12}$C,p)$^{23}$Na, have recently been studied at the surface of the earth \cite{Spillane07-PRL}. The off-resonance data were limited by the laboratory background \cite{Spillane07-PRL}. 
One can estimate the non-resonant contribution, to which any hypothetical resonance should be added, to be constant with energy and equal to the value found in Ref.~\cite{Spillane07-PRL}. Based on this assumption, for these two reactions a non-resonant cross-section about four orders below the present $\sigma_{\rm S=N}$ is found (table~\ref{table:Reactions}).  

For the present estimates (fig.~\ref{fig:minimumsigma}, table~\ref{table:Reactions}), the ion beam induced $\gamma$-background is assumed to be negligible at the $\gamma$-ray energies of interest. Whether or not this assumption is valid depends on the precise experimental setup, target, beam, and beam energy to be used. Therefore it is difficult to make generalized statements regarding the beam-induced background.

The present assumption of negligible beam induced $\gamma$-back\-ground was shown to be fulfilled in two selected cases with $\alpha$-beam at LUNA energies (section 4), and previously in one selected case for proton beam \cite{Bemmerer05-EPJA}. All of these cases involved gas targets. It is much more difficult to reliably predict the beam-induced background from $^{12}$C-beam at energies close to $E_{\rm Gamow}$, because unlike the proton- and $\alpha$-beam induced background this has not yet been investigated. In case a solid carbon target is selected, this presents an additional challenge. A dedicated study of $^{12}$C-beam induced $\gamma$-background at energies of 1-2\,MeV (much higher than the dynamic range of the current LUNA accelerator) is clearly called for as a first step to underground experiments with heavy ion beams such as $^{12}$C. 

In summary, for the present LUNA accelerator and present ultra-sensitive setup including a single large HPGe detector, it has been shown that cross sections of typically 1-10 pbarn can well be measured. The precise value depends on the $\gamma$-ray energy, and on the above discussed assumption that the beam induced background is under control. The extent to which the present feasibility data can be extended to proposed underground accelerator laboratories outside the LNGS facility \cite{Haxton07-NIMA,DOE08-arxiv,Strieder08-NPA3,Bordeanu08-NPA3} has to be evaluated based on the precise background conditions at those sites.

\section{Summary and outlook}

The feasibility of ultra-sensitive in-beam $\gamma$-ray spectroscopic studies for reactions with $\gamma$-rays of $E_\gamma$ $<$ 3\,MeV in the exit channel has been investigated at the LUNA deep underground accelerator facility.

To this end, the laboratory and ion-beam-induced $\gamma$-ray background for $\gamma$-energies $E_\gamma$ $<$ 3\,MeV has been studied. Using a sophisticated passive shielding, the laboratory $\gamma$-ray background for in-beam $\gamma$-spectroscopic studies has been reduced to levels not far from those achieved in state-of-the-art offline underground $\gamma$-counting. For two selected cases, the $\gamma$-ray background induced by an intensive $\alpha$-beam has been shown to be negligible when compared with the laboratory background. 

The data were then used to compute a $\gamma$-ray energy dependent cross section for which the expected signal is equal to the expected background. 

Based on several concrete cases, it has been shown that ultra-sensitive underground in-beam $\gamma$-ray spectroscopy has great potential for future contributions to experimental nuclear astrophysics.

\section*{Acknowledgments}

The present work has been supported by INFN and in part by the EU (ILIAS-TA RII3-CT-2004-506222), OTKA (T49245 and K68801), and DFG (Ro~429/41).


\begin{thebibliography}{10}

\bibitem{Greife94-NIMA}
U.~Greife {\em et~al.},
\newblock Nucl.~Inst.~Meth.~A {\bf 350}, 327 (1994).

\bibitem{MACRO90-PLB}
S.~P. {Ahlen} {\em et~al.},
\newblock Phys.~Lett.~B {\bf 249}, 149 (1990).

\bibitem{Belli89-NCA}
P.~Belli {\em et~al.},
\newblock Nuovo Cimento {\bf 101A}, 959 (1989).

\bibitem{NuPECC05-Roadmap}
{Nuclear Physics European Collaboration Committee (NuPECC), Roadmap 2005},
\newblock {available at http://www.nupecc.org/pub/NuPECC\_Roadmap.pdf}.

\bibitem{Haxton07-NIMA}
W.~Haxton, K.~Philpott, R.~Holtz, P.~Long, and J.~Wilkerson,
\newblock Nucl.~Inst.~Meth.~A {\bf 570}, 414 (2007).

\bibitem{DOE08-arxiv}
{{DOE/NSF Nuclear Science Advisory Committee}},
\newblock arXiV:0809.3137.

\bibitem{Strieder08-NPA3}
F.~Strieder,
\newblock J.~Phys.~G {\bf 35}, 014009 (2008).

\bibitem{Bordeanu08-NPA3}
C.~Bordeanu, C.~Rolfs, R.~Margineanu, F.~Negoita, and C.~Simion,
\newblock J.~Phys.~G {\bf 35}, 014011 (2008).

\bibitem{Arpesella96-Apradiso}
C.~Arpesella,
\newblock Appl. Radiat. Isot. {\bf 47}, 991 (1996).

\bibitem{Laubenstein04-Apradiso}
M.~Laubenstein {\em et~al.},
\newblock Appl.~Radiat.~Isot. {\bf 61}, 167 (2004).

\bibitem{Johansson95}
S.~Johansson, J.~Campbell, and K.~Malmqvist, editors,
\newblock {\em Particle-Induced X-Ray Emission Spectrometry} (Wiley, 1995).

\bibitem{Bemmerer05-EPJA}
D.~Bemmerer {\em et~al.},
\newblock Eur.~Phys.~J.~A {\bf 24}, 313 (2005).

\bibitem{Casella02-NPA}
C.~Casella {\em et~al.},
\newblock Nucl.~Phys.~A {\bf 706}, 203 (2002).

\bibitem{Bahcall05-ApJ}
J.~N. {Bahcall}, A.~M. {Serenelli}, and S.~{Basu},
\newblock Astrophys.~J. {\bf 621}, L85 (2005).

\bibitem{Formicola04-PLB}
A.~Formicola {\em et~al.},
\newblock Phys.~Lett.~B {\bf 591}, 61 (2004).

\bibitem{Imbriani05-EPJA}
G.~Imbriani {\em et~al.},
\newblock Eur.~Phys.~J.~A {\bf 25}, 455 (2005).

\bibitem{Lemut06-PLB}
A.~Lemut {\em et~al.},
\newblock Phys.~Lett.~B {\bf 634}, 483 (2006).

\bibitem{Bemmerer06-NPA}
D.~Bemmerer {\em et~al.},
\newblock Nucl.~Phys.~A {\bf 779}, 297 (2006).

\bibitem{Marta08-PRC}
M.~{Marta} {\em et~al.},
\newblock Phys.~Rev.~C {\bf 78}, 022802(R) (2008).

\bibitem{Haxton08-ApJ}
W.~C. {Haxton} and A.~M. {Serenelli},
\newblock Astrophys.~J. {\bf 687}, 678 (2008).

\bibitem{Imbriani04-AA}
G.~Imbriani {\em et~al.},
\newblock Astron.~Astrophys. {\bf 420}, 625 (2004).

\bibitem{Herwig04-ApJ}
F.~{Herwig} and S.~M. {Austin},
\newblock Astrophys.~J. {\bf 613}, L73 (2004).

\bibitem{Herwig06-PRC}
F.~{Herwig}, S.~M. {Austin}, and J.~C. {Lattanzio},
\newblock Phys.~Rev.~C {\bf 73}, 025802 (2006).

\bibitem{Formicola08-NPA3}
A.~Formicola {\em et~al.},
\newblock J.~Phys.~G {\bf 35}, 014013 (2008).

\bibitem{Diehl06-Nature}
R.~Diehl {\em et~al.},
\newblock Nature {\bf 439}, 45 (2006).

\bibitem{Iliadis02-ApJSS}
C.~Iliadis, A.~Champagne, J.~Jos{\'e}, S.~Starrfield, and P.~Tupper,
\newblock Astrophys.~J.~Suppl.~Ser. {\bf 142}, 105 (2002).

\bibitem{Bemmerer06-PRL}
D.~{Bemmerer} {\em et~al.},
\newblock Phys.~Rev.~Lett. {\bf 97}, 122502 (2006).

\bibitem{Gyurky07-PRC}
G.~{Gy{\"u}rky} {\em et~al.},
\newblock Phys.~Rev.~C {\bf 75}, 035805 (2007).

\bibitem{Confortola07-PRC}
F.~{Confortola} {\em et~al.},
\newblock Phys.~Rev.~C {\bf 75}, 065803 (2007).

\bibitem{Costantini08-NPA}
H.~Costantini {\em et~al.},
\newblock Nucl.~Phys.~A {\bf 814}, 144 (2008).

\bibitem{Serpico04-JCAP}
P.~D. Serpico {\em et~al.},
\newblock Journal of Cosmology and Astroparticle Physics {\bf 2004}, 010
  (2004).

\bibitem{Rolfs74-NPA-12C}
C.~Rolfs and R.~Azuma,
\newblock Nucl.~Phys.~A {\bf 227}, 291 (1974).

\bibitem{Spillane07-PRL}
T.~Spillane {\em et~al.},
\newblock Phys.~Rev.~Lett. {\bf 98}, 122501 (2007).

\bibitem{Iliadis07}
C.~Iliadis,
\newblock {\em Nuclear Physics of Stars} (Wiley-VCH, 2007).

\bibitem{Powell99-NPA}
D.~C. Powell {\em et~al.},
\newblock Nucl.~Phys.~A {\bf 660}, 349 (1999).

\bibitem{Formicola03-NIMA}
A.~Formicola {\em et~al.},
\newblock Nucl.~Inst.~Meth.~A {\bf 507}, 609 (2003).

\bibitem{Casella02-NIMA}
C.~Casella {\em et~al.},
\newblock Nucl.~Inst.~Meth.~A {\bf 489}, 160 (2002).

\bibitem{Marta06-NIMA}
M.~Marta {\em et~al.},
\newblock Nucl.~Inst.~Meth.~A {\bf 569} (2006).

\bibitem{Heusser93-NIMB}
G.~Heusser,
\newblock Nucl.~Inst.~Meth.~B {\bf 83}, 223 (1993).

\end{thebibliography}

\end{document}